\documentstyle[espcrc2,fleqn,twoside,epsfig]{article}

\begin{document}

\title{Thermodynamics and tunneling spectroscopy
       in the pseudogap regime \\
       of the boson fermion model}
 
\author{T. Doma\'nski
       \address[IFUMCS]
          {Institute of Physics, Maria Curie Sk\l odowska
           University, 20-031 Lublin, Poland}
       \address[CRTBT]
          {Centre de Recherches sur les Tr\`es Basses
          Temp\'eratures CNRS, 38-042 Grenoble Cedex 9, France}
       and J. Ranninger
       \addressmark[CRTBT]
       }

\begin{abstract}
Motivated by the STM experimental data on Bi$_{2}$Sr$_{2}$CaCU$_{2}$O$_{8+x}$
which indicate the tunneling conductance asymmetry $\sigma(-V)\neq\sigma(V)$,
we report that such a behavior can be explained in terms of the boson
fermion model. It has been shown in the recent studies, based on various
selfconsistent techniques to capture the many-body effects, that the low
energy spectrum of the boson fermion model is featured by an appearance
of the pseudogap at $T^{*}>T_{c}$. We argue that the pseudogap structure
has to exhibit a particle-hole asymmetry. This asymmetry may eventually
depend on the boson concentration.
\end{abstract}

\maketitle

\section{Introduction}

Recently there was a considerable amount of studies of the 
pseudogap phenomenon observed in a variety of experiments on 
the high temperature superconductors (HTSC) \cite{Timusk-99}. 
There are two main theoretical interpretations which are 
presently widely considered in the literature:
 (i) the pseudogap as a precursor of the emerging 
     pairing fluctuations, and 
(ii) the pseudogap understood in terms of some new (hidden) 
     ordering taking place in a vicinity of the superconducting 
     phase.
Some selective overview can be found e.g.\ in the recent
monograph \cite{Carlson-02}.

Among theoretical attempts to explain the pseudogap effect
of HTSC materials there is a model of itinerant electrons
or holes which coexist and interact with the local bound 
pairs (hard-core bosons) \cite{Ranninger-85}. It is worth
mentioning, that the pseudogap has been foreseen within 
this boson fermion (BF) model a long time before the convincing 
experimental data became available (see the last paragraph 
in section IV of the Ref.\ \cite{Micnas-90}).

Pseudogap phase of the BF model is a manifestation of the 
pairing-wise correlations which start to appear in a system 
when the transition temperature $T_{c}$ is approached from 
above (the precursor type interpretation). On a microscopic 
basis it means that below a certain temperature $T^{*}$
fermions start to couple into the pairs. These are, however, 
weakly ordered in phase due to the small superfluid stiffness.
The phase coherence sets in at a sufficiently low temperature 
$T_{c} \leq T^{*}$ and then the true superconducting transition 
occurs. Presence of the incoherent ($T^{*}<T<T_{c}$) or the 
coherent ($T_{c} \geq T$) fermion pairs is accompanied by  
either the pseudogap or the true superconducting gap formed 
around the Fermi energy $\varepsilon_{F}$.

In general, some advanced methods of the many body theory are 
required to explore a pseudogap phase within any model. It is
because the single-particle and the two-particle correlations 
are then of equal importance. They should be properly treated 
taking account of possible feedback effects between both channels 
in a controlled way. In a context of the BF model such requirements 
were obtained so far via: a) the selfconsistent perturbative 
investigation \cite{perturbative,Ren-98}, b) the dynamical mean field 
theory \cite{DMFT}, and c) the flow equation study \cite{Domanski-01}. 
Some alternative way was studied by Micnas {\em et al} 
\cite{Micnas-01} who considered the Kosterlitz-Thouless criterion 
for determination of the superconducting transition $T_{c}=T_{KT}$ 
and identified $T^{*}$ with the mean field estimate $T_{c}^{(MF)}$.
Authors were able to reproduce qualitatively the Uemura type 
plots $T_{c}$ versus $\rho_{s}$.

In this short paper we extend our previous study \cite{Domanski-01}
to analyze the pseudogap shape and its variation with temperature. 
Direct consequences of the particle-hole asymmetric pseudogap 
are illustrated on an example of the STM current conductance.

\section{Model and the origin of correlations} 

We consider the BF model which is described by the following
Hamiltonian \cite{Ranninger-85,Micnas-90}
\begin{eqnarray}
H^{BF} & = & \sum_{i,j,\sigma} \left( t_{ij} - \mu
\delta_{i,j} \right) c_{i\sigma}^{\dagger} c_{j\sigma}
+ \sum_{i} E_{0} b_{i}^{\dagger} b_{i}
\nonumber \\ & + &  
v \sum_{i} \left(  b_{i}^{\dagger} c_{i\downarrow}
c_{i\uparrow} +  b_{i} c_{i\uparrow}^{\dagger}
c_{i\downarrow}^{\dagger} \right) 
\label{BF}
\end{eqnarray}
where $E_{0}=\Delta_{B}-2\mu$ and $\Delta_{B}$ is the boson energy,
$\mu$ is the chemical potential. The second quantization operators 
$c_{i\sigma}^{(\dagger)}$ refer to the conduction band particles
(electrons or holes) and $b_{i}^{(\dagger)}$ to the composite hard-core 
bosons (for instance they can represent the trapped electron pairs 
$b_{i}=d_{i\downarrow}d_{i\uparrow}$). Itinerant fermions propagate 
between the lattice sites $i$ and $j$ via the hopping integral $t_{ij}$ 
whereas bosons are assumed to be infinitely heavy. The hard-core 
property of bosons means that either $0$ or $1$ boson is allowed 
to occupy a given lattice site. This local constraint can be formally 
expressed \cite{Micnas-90} through the following semi-bosonic commutation 
relations $\left[ b_{i},b_{j}^{\dagger}\right] = \delta_{i,j}
\left(1-2b_{i}^{\dagger}b_{i}\right)$ and $\left[b_{i},b_{j}\right]
=0=\left[b_{i}^{\dagger},b_{j}^{\dagger}\right]$. 

Mechanism of superconductivity and all the other forms of 
correlations in the BF model (\ref{BF}) are caused by the boson 
fermion charge exchange potential $v$. By decaying into the fermion 
pairs, bosons gain effectively some mobility. If temperature  
decreases below the critical value $T_{c}$ then (for $dim>2$) 
some fraction of bosons gets "frozen" into the BE condensate 
$n_{0}(T) = \frac{1}{N} \left< b_{{\bf q}={\bf 0}}^{\dagger}
b_{{\bf q}={\bf 0}} \right>$. For $n_{0}(T)\neq 0$, fermions 
are simultaneously driven into the broken symmetry superconducting 
state. It can be shown \cite{Micnas-90,Domanski-01} that 
the energy gap in the superconducting fermion subsystem 
is $v \sqrt{n_{0}(T)}$.

For temperatures slightly higher than $T_{c}$ there exist 
many bosons which occupy the small momenta ${\bf q} \sim {\bf 0}$ 
states. Because of the interaction $v \sum_{{\bf k},{\bf q}} 
\left( b_{\bf q}^{\dagger} c_{{\bf k}+{\bf q}/2\downarrow}
c_{-{\bf k}+{\bf q}/2\uparrow} + h.c. \right)$, these boson 
states $\left| n_{\bf q} \right>_{B}$ are strongly mixed with 
the fermion states $\left| n_{{\bf k}+{\bf q}/2\downarrow}
\right>_{F}$ $\left| n_{-{\bf k}+{\bf q}/2\uparrow} \right>_{F}$  
and thereby the life time of fermions might be reduced, 
especially for $|{\bf k}| \sim k_{F}$. In consequence,
we expect that the fermion density of states might be
suppressed near $\varepsilon_{F}$.

With the on-site boson fermion interaction given in (\ref{BF}) 
one can generate only the isotropic gap/pseudogap.  Of course, 
the HTSC materials are characterized by the anisotropic order 
parameters of the $d$-wave symmetry with a possible admixture 
of the $s$-wave component \cite{mixed_symmetry}. To capture
this aspect it is enough to introduce the intersite coupling 
$v_{i,j}  b_{i}^{\dagger} \left( c_{i\downarrow} c_{j\uparrow} 
+ c_{j\downarrow}c_{i\uparrow} \right) + h.c.$ when both, the 
superconducting gap \cite{Micnas-01,Domanski-02,anisotropicBF} 
and the pseudogap \cite{Ren-98} become anisotropic.  Here we 
only discuss the results for the isotropic case but, at a price 
of more difficult numerical computations, the same procedure 
can be easily extended to the anisotropic pairing.
 
\section{The effective spectra}

In order to determine the effective fermion and boson spectra 
of the model (\ref{BF}) we utilize the flow equation technique 
proposed by Wegner \cite{Wegner-94}. The main idea behind is 
to disentangle the coupled boson and fermion subsystems via 
a sequence of canonical transformations $H(l)=e^{-S(l)}He^{S(l)}$, 
where $l$ is a continuous parameter. We start at $l=0$    
by putting $H^{BF} \equiv H(0)$, and proceed till $l=\infty$,
when we want to obtain $H(\infty)=H^{F}_{eff}+H^{B}_{eff}$. 
All the way, from $l=0$ to $l=\infty$, we adjust the operator 
$S(l)$ according to the Wegner's prescription \cite{Wegner-94}.
In practice, disentangling of fermion from boson subsystem can be 
done within an accuracy of the order $v^{3}$ \cite{Domanski-01}.
To simplify the matters we neglect the hard-core constraint 
and use the pure bosonic relations $\left[ b_{i},b_{j}^{\dagger} 
\right] \simeq \delta_{i,j}$ which should be valid for small 
boson concentrations $n_{B}=\left< b_{i}^{\dagger} b_{i}\right>$. 

After the disentangling procedure is finished we obtain 
the following structure for the boson contribution to
the effective Hamiltonian $H^{B}_{eff}=\sum_{\bf q} 
\left( E_{\bf q}-2\mu\right)$. The initial boson energy
$\Delta_{B}$ is thus transformed into the dispersion
$E_{\bf q}$ which is characterized by the width proportional
to $v^{2}$ and the effective boson mass comparable with 
the mass of fermions \cite{perturbative,Domanski-01}. 
$H^{F}_{eff}$ part, on the other hand, is given as
$H^{F}_{eff}=\sum_{{\bf k},\sigma} \left(\varepsilon_{\bf k}-\mu
\right) c_{{\bf k}\sigma}^{\dagger}c_{{\bf k}\sigma} + \frac{1}{N}
\sum_{{\bf k},{\bf p},{\bf q}} U_{{\bf k},{\bf p},{\bf q}} 
c_{{\bf k}\uparrow}^{\dagger} c_{{\bf p}\downarrow}^{\dagger} 
c_{{\bf q}\downarrow}c_{{\bf k}+{\bf p}-{\bf q}\uparrow}$.
Renormalization of $\varepsilon_{\bf k}$ with respect to the
initial dispersion $\varepsilon_{\bf k}^{0}$ takes place 
mainly around ${\bf k}_{F}$. There is also induced the long range 
fermion-fermion interaction $U_{{\bf k},{\bf p},{\bf q}}$ 
which has somewhat unusual resonant-like character as shown
in Figs 7 and 8 of \cite{Domanski-01} for the BCS 
$U_{{\bf k},-{\bf k},{\bf q}}$ and for the density-density 
$U_{{\bf k},{\bf q},{\bf q}}$ channels.

Previously \cite{Domanski-01} we discussed the fermion
spectrum only on a basis of the quasiparticle energy 
$\varepsilon_{\bf k}$. However, in some cases,
a considerable influence may also arise from the 
fermion-fermion interactions. These interactions are in 
principle small, $|U_{{\bf k},{\bf p},{\bf q}}|$ is of 
the order $v^{2}$, so we can treat them perturbatively. 
The effective dispersion $\overline{\varepsilon}_{{\bf k}\sigma}(l)$  
is for $T>T_{c}$ given by
\begin{eqnarray}
\overline{\varepsilon}_{{\bf k}\uparrow} = \varepsilon_{\bf k}
+ \frac{1}{N}\sum_{\bf q} U_{{\bf k},{\bf q},{\bf q}} 
\left< c_{{\bf q}\downarrow}^{\dagger} c_{{\bf q}\downarrow} 
\right> .
\label{correction}
\end{eqnarray}
At $T<T_{c}$ one should also consider the other contribution 
coming from the BCS channel $U_{{\bf k},-{\bf k},{\bf q}}$. 
We restrict our attention only to the normal phase ($T>T_{c}$).

\begin{figure}
\centerline{\epsfig{file=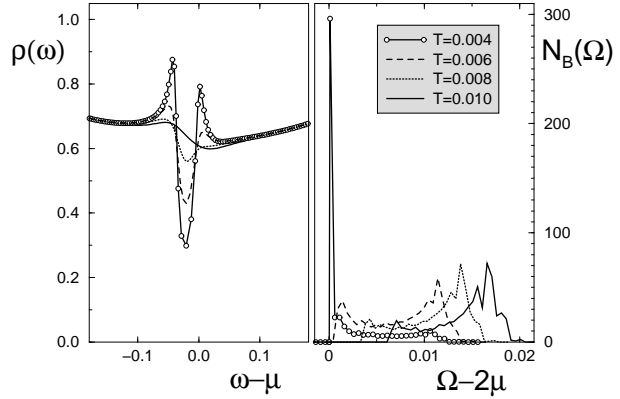, width=8cm}}
\caption{The density of fermion states $\rho(\omega)$ and 
the boson occupancy $N_{B}(\Omega)$ of the BF model with 
$v=0.1$, $\Delta_{B}=0$ and concentrations $n_{F}=1$, 
$n_{B}=0.3$. Energies $\omega$, $\Omega$, $\mu$, $v$ 
and $k_{B}T$ are all expressed in units of the femion bandwidth 
$D$.}
\label{figure1}
\end{figure}

In the left h.s.\ panel of figure \ref{figure1} 
we show the density of fermion states $\rho(\omega) \equiv 
\frac{1}{N}\sum_{\bf k} \delta \left( \omega - \overline
{\varepsilon}_{\bf k} \right)$. Note, that there appears
a pseudogap which deepens with a decreasing temperature $T$. 
The pseudogap structure has a clear particle-hole asymmetry
at all the temperatures. Asymmetry finally disappears 
at very high temperatures $T \simeq 0.1$ (not shown here).
The boson density of states $\frac{1}{N} \sum_{\bf q} \delta 
\left( \Omega - E_{\bf q} \right)$ is much less affected by 
a varying temperature (see Fig.\ 4 in \cite{Domanski-01}).
However, upon decreasing $T$ we observe (see the right h.s.\
panel of figure \ref{figure1}) a considerable redistribution 
of boson occupancy $N_{B}(\Omega)=\frac{1}{N} \sum_{\bf q} 
\delta \left(\Omega - E_{\bf q} \right) f_{BE} \left(E_{\bf q}
-2\mu,T\right)$, here $f_{BE}$ is the Bose Einstein distribution.
By comparing both the panels of figure \ref{figure1} we notice
that the pseudogap builds up when bosons start populating 
the low energy states  $E_{{\bf q} \simeq {\bf 0}}$. 
%

Asymmetry of the pseudogap structure is mainly controlled 
by the boson concentration $n_{B}$. Here is a simple 
argumentation. If the boson energy is located in a center 
of the fermion band ($\Delta_{B}=0$), then for the exactly 
half-filled fermion and boson subsystems they both must 
have symmetric spectra. In particular, the pseudogap would 
then become symmetric too. For the situation presented 
in Fig.\ \ref{figure1} we have $n_{F} \simeq 1$, so it 
can only be the boson concentration $n_{B}$ responsible 
for the asymmetry of $\rho(\omega)$. A more detailed 
analysis will be presented in the future publication.

\section{Single particle spectroscopy} 

Our results, in particular effects of the particle-hole
asymmetry, can be well illustrated by calculating
the single particle tunneling current $J$. The differential 
conductance $\sigma(V)=dJ/dV$ as a function of bias 
voltage $V$ is a direct probe of the density of states
below and above the Fermi energy. We use the following
expression for the STM current
\begin{eqnarray}
J(V) = const \int_{-\infty}^{\infty} d \omega \; \rho(\omega) 
\; \left[ f_{FD}(\omega,T) \right.
\nonumber \\ 
\left. - f_{FD}(\omega-eV,T) \right] ,
\label{J}
\end{eqnarray}
where $f_{FD}$ stands for the Fermi Dirac distribution.
As usually, we neglect the energy $\omega$ and 
${\bf k}$-dependence of the tunneling matrix 
\cite{Norman-00}.

\begin{figure}
\centerline{\epsfig{file=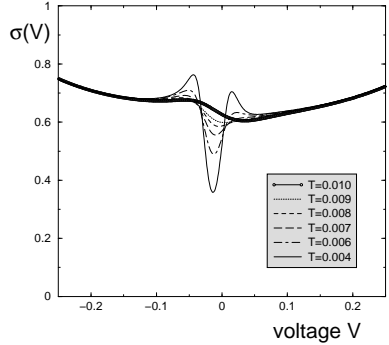, width=5cm}}
\caption{
The STM current conductance $\sigma(V)$ as a function
of bias voltage $V$ (in units of $D/e$) for the same
set of parameters as shown in Fig.\ \ref{figure1}.}
\label{figure2}
\end{figure}

In figure \ref{figure2} we show the conductance $\sigma(V)$ 
of the STM current (\ref{J}) obtained for the same set of 
parameters as in Fig.\ \ref{figure1}. We obtain 
the negative-positive asymmetric characteristics because, 
at low T, the conductance is roughly proportional to the 
density of states $\rho(\omega)$. Our results agree very 
well with the experimental data reported by Renner {\em et al} 
\cite{Renner-96}. Unfortunately, we are unable to pass 
through $T_{c}$ (we solve the flow equations using the one 
dimensional tight binding dispersion $\varepsilon_{\bf k}^{0}$ 
\cite{Domanski-01} when $T_{c}^{dim=1}=0$). However,
for the realistic $dim>2$, we expect the asymmetry to 
survive even at $T<T_{c}$ as seen experimentally 
\cite{Renner-96}.
%

Let us point out the main features of the pseudogap probed 
experimentally by the STM conductance \cite{Renner-96}: 
  (i) it is asymmetric, 
 (ii) its magnitude (the peak to peak distance) 
      is almost temperature independent, and 
(iii) the pseudogap deepens with a decreasing temperature
      while the coherence peaks gradually start to appear. 
The BF model is capable to reproduce all these features 
(i)-(iii). Some other theoretical concepts discussed in 
the literature to explain $\sigma(V)$, e.g.\ \cite{Norman-00} 
and references cited therein, are dealing with the physics
which microscopically is very close to the BF model (\ref{BF}).

{\bf Acknowledgment} 
T.D.\ kindly acknowledges hospitality of the J.\ Fourier 
University and Centre de Recherches sur les Tr\`es Basses 
Temperatures in Grenoble, where this study was carried out. 
The work was partly supported by the Polish State Committee 
for Scientific Research, grant No.\ 2P03B 106 18.

\end{document}